\documentclass[12pt]{amsart}
\usepackage{amssymb}
\usepackage{verbatim}
\usepackage[usenames]{color}
\usepackage{hyperref}

\newtheorem{thm}{Theorem}

\theoremstyle{definition}

\theoremstyle{remark}

\newtheorem{rmk}[thm]{Remark}

\newenvironment{ls}{\begin{itemize}}{\end{itemize}}
\newenvironment{lsnum}{\begin{enumerate}}{\end{enumerate}}

\newcommand{\ger}[1]{\ensuremath{\mathfrak {#1}}}

\newcommand{\bbb}[1]{\ensuremath{\mathbb {#1}}}
\newcommand{\ttt}[1]{\ensuremath{\mathtt {#1}}}

\newcommand{\emp}{\varnothing}
\renewcommand{\phi}{\varphi}

\newcommand{\sq}[1]{\ensuremath{\langle#1\rangle}}

\newcommand{\notarrow}{\kern .42em\not\kern -.42em\longrightarrow}

\newcommand{\noprint}[1]{\relax}

\newcommand{\q}{\medskip\noindent\textbf{Q:\ }}
\renewcommand{\a}{\medskip\noindent\textbf{A:\ }}

\title[Two forms of one useful logic]%
{Two Forms of One Useful Logic:\\
Existential Fixed Point Logic and\\
Liberal Datalog}
\author{Andreas Blass}
\address{Mathematics Department\\
University of Michigan\\
Ann Arbor, MI 48109--1043, U.S.A.}
\email{ablass@umich.edu}
\thanks{Partially supported by NSF grant DMS-0653696}
\author{Yuri Gurevich}
\address{Microsoft Research\\
  One Microsoft Way\\
  Redmond, WA \ 98052, U.S.A.}
\email{gurevich@microsoft.com}

\begin{document}
\maketitle

\begin{abstract}
  A natural liberalization of Datalog is used in the Distributed
  Knowledge Authorization Language (DKAL).  We show that the
  expressive power of this liberal Datalog is that of existential
  fixed-point logic. The exposition is self-contained.
\end{abstract}

\section{Prologue}

Existential fixed point logic (EFPL) differs from first-order
logic by prohibiting universal quantification (while allowing
existential quantification) and by allowing the ``least fixed point''
operator for positive inductive definitions.  A precise definition is
given below.

Our original motivation for developing EFPL in \cite{efp} was its
appropriateness for formulating pre- and post-conditions in Hoare's
logic of asserted programs \cite{hoare}.  In particular, the
expressivity hypothesis needed for Cook's completeness theorem
\cite{cook} in the context of first-order logic is automatically
satisfied in the context of EFPL.

But it turned out that EFPL has many other interesting properties.
\begin{lsnum}
\item EFPL captures polynomial time computability on the class of
  structures of the form $\{0,1,\dots,n\}$ with (at least) the
  successor relation and names for the endpoints.
\item The set of logically valid EFPL formulas is a complete
  recursively enumerable set.
\item The set of satisfiable EFPL formulas is a complete
  recursively enumerable set.
\item The set of EFPL formulas that hold in all finite structures is a
  complete co-r.e. set.
\item When an EFPL formula is satisfied by a tuple of elements in a
  structure, this fact depends on only a finite part of the structure.
\item No transfinite iteration is needed when evaluating EFPL
  formulas, by the natural iterative process, in any
  structure.\footnote{This means that the closure ordinal of each of
    the iterations is at most $\omega$, the first infinite ordinal.
    Contrast this with what happens when the least fixed point
    operator is added to full first-order logic or just to its
    universal fragment.  As shown in the appendix to \cite{efp},
    arbitrarily large closure ordinals are possible there.}
\item EFPL can define (given appropriate syntactic apparatus) truth of
  EFPL formulas.
\item Truth of EFPL formulas is preserved by homomorphisms.
\item If an EFPL formula and a first-order formula are equivalent,
  then they are equivalent to an existential\footnote{We could nearly
    say ``existential positive'' here.  Negative occurrences are
    needed in the existential formula only for those predicate symbols
    that are negative in the vocabulary of the EFPL formula.}
  first-order formula.
\end{lsnum}
Except for (7), which will be proved elsewhere, all these results are
in \cite{efp}.  The combination of (2) and (3) is surprising; it is
possible because EFPL is not closed under negation.  That is also why
(7) doesn't contradict Tarski's theorem on undefinability of truth.

Recently, EFPL has found a new application as the logical underpinning
of the distributed knowledge authorization language DKAL \cite{dkal,
  dkal-abs}.  For this application, it was useful to recast EFPL in a
form that looks similar to Datalog; it was called liberal Datalog in
\cite{dkal-abs}.  The purpose of the present note is to show exactly
how the logic programs of liberal Datalog correspond to the formulas
of traditional EFPL.

\section{Introduction}

\noindent\textbf{Quisani:\ }
Let's return to existential fixed-point logic. We discussed it once
\cite{underl}, yet something bothers me about the definition.

\noindent\textbf{Authors:}\footnote{Not necessarily speaking in
    unison.}\
Before we get to what's bothering you, let's be sure you have the
correct definition from \cite{efp}.

\q I think I know the definition all right, but to be safe let me
check it with you: After making the convention that predicate symbols
are classified as \emph{positive} or \emph{negatable}, one defines
terms and atomic formulas just as in first-order logic.  Compound
formulas are built by
\begin{ls}
  \item negation, applied only to atomic formulas whose predicate
    symbol is negatable,
\item conjunction and disjunction,
\item existential quantification, and
\item the LET-THEN construction.
\end{ls}
All but the last of these have their traditional meanings as in
first-order logic.  The LET-THEN construction\footnote{Other notations
are ``let \dots\ in \dots\ ''and ``letrec \dots\ in \dots\ .''  We
retain ``then'' mainly for consistency with \cite{efp}.  ``Then'' also
serves as a reminder that, when expressed in second-order logic, the
contruction amounts to an implication: ``If you interpret the $P_i$'s
in such a way that each $P_i$ is implied by the corresponding
$\delta_i$, \textbf{then} $\psi$ holds.''} produces formulas of the
form
\[
\text{LET }P_1(\vec x^1)\leftarrow\delta_1,\dots,P_k(\vec
x^k)\leftarrow\delta_k\text{ THEN }\psi
\]
where the $P_i$'s are distinct, new, positive predicate symbols and
the $\delta_i$ and $\psi$ are EFPL formulas in the vocabulary expanded
by addition of these $P_i$'s.  Semantically, this formula means to use
the $\delta_i$'s to define a monotone operator on $k$-tuples of
predicates; given a tuple of (interpretations of) the $P_i$'s, see
which tuples $\vec x^i$ satisfy the $\delta_i$'s, and use those sets
of tuples as the new interpretation of the $P_i$'s.  Repeat this
operation until you reach a fixed point.  Finally, use this
least fixed point of the operator to interpret the $P_i$'s in
$\psi$.

\a
That's right. You tacitly assumed a specific vocabulary $\Upsilon$
when you said that the $P_i$'s are new, meaning they're not in
$\Upsilon$.  In other words, although they occur in the LET-THEN
formula you mentioned, they don't count as part of the vocabulary of
that formula.

\q
Right. I think of them as bound predicate variables.  They could, for
example, be renamed without affecting the meaning of the formula (as
long as there are no clashes).

\a Indeed, bound predicate variables are just what the $P_i$'s become
when the formula is translated into second-order logic (as in
Theorem~5 of \cite{efp}).  That reminds us of another comment about
your description of the semantics.  You repeatedly applied the operator
defined by the $\delta_i$'s, until a fixed point is reached.  But the
semantics merely requires the least fixed point; it doesn't care about
its explicit construction.

\q
OK.  I guess I was giving a sort of operational semantics, whereas the
logic is defined in a purely denotational way.

But I've found it convenient to think about EFPL operationally,
especially when trying to write EFPL formulas to express particular
properties.  For example, I once checked that, in the standard model
of arithmetic, $\ger N=\sq{N,0,1,+,\cdot,<}$, the property of being a
prime number is expressible by an EFPL formula.

\a By Matiyasevich's solution of Hilbert's tenth problem, this
property --- indeed any recursively enumerable property of natural
numbers --- is expressible in \ger N by an existential first-order
formula; you don't need the fixed point operator.

\q I know, but I was looking for a formula that directly expresses
what it means to be prime, without detouring through clever
Diophantine tricks.  The formula I constructed was actually fairly
complicated, mainly because of the need to simulate two bounded
universal quantifiers.  A natural definition of ``$x$ is prime'' is
that no $u < x$ divides $x$ unless $u = 1$, and a natural definition
of ``$u$ doesn't divide $x$'' (as long as $1\neq u<x$) is that there
is no $v < x$ such that $u \cdot v = x$.  These are the two bounded
universal quantifications.  In \cite{efp}, you showed\footnote{See the
  proof of Theorem~3 in \cite{efp}.} how to replace some cases of
universal quantification with EFPL descriptions of searches through
the domains of quantification. For example, in the case of arithmetic,
$(\forall w < x) P(w)$ can be replaced with
\[
\text{LET }Y(u)\leftarrow u=0\lor\exists w\,[u=w+1\land Y(w)\land
P(w)] \text{ THEN }Y(x).
\]
I applied this idea, replacing each of the two universal quantifiers
in the definition of ``prime'' by a search.  Here's what I came up
with, assuming that equality is negatable.
\begin{align*}
  &\text{LET }Y(u,v)\leftarrow\\
&\qquad u=0\lor
\exists w\,[u=w+1\land Y(w,v)\land(w\cdot v\neq x\lor w=1\lor w=x)
\\ &\text{THEN }\\
&\qquad\text{LET }X(u,v)\leftarrow \\
&\qquad\qquad v=0\lor\exists w\,[v=w+1\land
X(u,w)\land Y(u,w)]\\
&\qquad\text{THEN }
1<x\land X(x,x).
\end{align*}

\a That looks correct.  The first LET-clause makes $Y(u,v)$ express
``$x$ is not the product of anything $<u$ with $v$, except for trivial
products $1\cdot x$ and $x\cdot 1$,'' and then the second LET-clause
makes $X(u,v)$ express ``$x$ is not a nontrivial product of anything
$<u$ and anything $<v$.''  So, as you said, the two LET-clauses
replace the two universal quantifiers in the natural first-order
definition of ``prime.''  By the way, you don't really need that
equality is negatable, since you can replace $w\cdot v\neq x$ with
$(w\cdot v<x)\lor(x<w\cdot v)$.

\q That's right.  And I don't really need $<$ since $a<b$ is
equivalent to $\exists y\,(a+y+1=b)$.  But I wasn't trying to minimize
the vocabulary; I just wanted to make sure I see how to formalize
things in EFPL.

\a OK.  Did you notice that, although your formulas define the desired
predicates on all of \bbb N, you could have cut off the searches at
$x$, for example by adding a conjunct $w<x$ for each $\exists w$?  The
resulting finite searches would still define ``prime'' correctly.

\q I thought of that, but I decided the formula was long enough
already.

\a In any case, it's clear that you know what EFPL is; so what's the
problem with the definition?

\q I thought I knew EFPL until I saw your extended abstract about the
distributed knowledge authorization language, DKAL \cite{dkal-abs}.

\a That abstract isn't by the two of us; it's by one of us and Itay
Neeman.

\q I know, but Itay isn't here just now and you are, so I hope you can
clarify the situation for me.

\a We'll try.  What exactly needs to be clarified?

\q Section~2 of the extended abstract claims to be about existential
fixed-point logic (EFPL), but it looks quite different from the logic
that I learned from your paper \cite{efp} and described to you here.
In particular, that
section of the abstract hardly mentions quantifiers at all and makes
no distinction between existential and universal quantifiers, whereas
that distinction was crucial in \cite{efp}.

So I decided to look at the full tech report \cite{dkal}.  Its
Section~2 is very similar to that of the extended abstract.  Its
Appendix~A.3 contains a quick, prose description of EFPL as defined in
\cite{efp} but then ignores that and talks about logic programs and
queries instead, just as Section~2 did.

As a result, I'm wondering about the connection between the ``logic
programs plus queries'' picture in \cite{dkal-abs, dkal} and the
traditional picture of EFPL in \cite{efp}.

\a The traditional picture in \cite{efp} corresponds exactly to the
logic programs aspect described in the DKAL paper \cite{dkal}.  The
queries in the latter paper are outside EFPL, because they include
universal quantification, at least in certain circumstances.

When only relational structures are considered, so that logic programs
amount to Datalog, their equivalence with EFPL is in \cite{ch}.
Grohe mentioned in the introduction of \cite{grohe} that it extends to
the case of vocabularies that include function symbols.

\q Is the general case proved there?  If not, can you show me in
detail how logic programs correspond to traditional EFPL formulas?

\a We don't recall seeing a published source for the details of the
correspondence in the general case.  First, let's state the
correspondence precisely.  We deal with structures $X$ for a
vocabulary in which all predicate symbols are negatable.  (So positive
predicate symbols will arise only as the $P$'s in LET-clauses.)

\begin{thm}
  The relations definable, in a structure $X$, by EFPL formulas
  are the same as the superstrate relations obtained, over the
  substrate $X$, by logic programs.
\end{thm}

The proof is in two parts, namely translations in both
directions between the two formalisms.  Furthermore, the translations
are uniform; that is, they do not depend on the structure $X$.
Incorporating this uniformity into the statement of the theorem, we
have the following more complete formulation.

\begin{thm}
  For every EFPL formula $\phi$ with free variables among
  $x_1,\dots,x_n$, there is a logic program $\Pi$ with a distinguished
  $n$-ary superstrate relation $P$ such that, in every structure $X$,
  the interpretation of $P$ defined by $\Pi$ consists of exactly the
  $n$-tuples that satisfy $\phi$.  Conversely, given a logic program
  $\Pi$ and a distinguished superstrate predicate symbol $P$, there is
  an EFPL formula $\phi$ defining, in every structure $X$, the set of
  $n$-tuples that $\Pi$ produces as the interpretation of $P$.
\end{thm}

\q
You said ``every structure $X$'' but surely you intended some restriction
on the vocabulary of $X$.

\a You're right.  The vocabulary of $X$ should the same as that of
$\phi$. That's also the substrate vocabulary of $\Pi$, while the full
vocabulary of the program $\Pi$ includes, in addition, $P$ and
(possibly) other superstrate predicate symbols.

\section{From Logic Programs to Formulas}

\a Let's begin by considering a logic program of the sort described in
\cite{dkal}.  To recapitulate that description, we have a vocabulary
$\Upsilon$ divided into a \emph{substrate} part (which may contain
relation and function symbols) and a \emph{superstrate} part
(containing only relation symbols).  Substrate (resp.\ superstrate)
formulas are those whose relation symbols are all in the substrate
(resp.\ superstrate) part of $\Upsilon$; note that a superstrate
formula is allowed to contain substrate function symbols.

\q What's the intuition behind substrate and superstrate?

\a The idea is that the substrate relations are given to us and the
superstrate relations are computed by means of the program.  That
intuition is reflected in the semantics, which we'll review in a
moment, but first let's finish the description of the syntax of logic
programs.

A \emph{logic rule} in the vocabulary $\Upsilon$ has the form
$H\leftarrow B$, where the \emph{head} $H$ is an atomic superstrate
formula and the \emph{body} $B$ is a conjunction of atomic superstrate
formulas and possibly a quantifier-free substrate formula.  A
\emph{logic program} is a finite set of logic rules.

Semantically, a logic program is to be interpreted in a given
structure $X$ for the substrate vocabulary.  It defines
interpretations for the superstrate relations as the least fixed point
of all the rules in the program.  That is, interpret each rule
\[
R(t_1,\dots,t_n)\leftarrow B
\]
in the program as an instruction to increase the current
interpretation of $R$ by adding all those tuples $(a_1,\dots,a_n)$ of
elements of $X$ such that some assignment of values (in $X$) to the
variables makes $B$ true and gives each $t_i$ the value $a_i$.
Formally, this amounts to an operator $\Gamma$ on tuples of relations
regarded as interpreting all the superstrate relations (or,
equivalently, on $\Upsilon$-structures whose reduct to the substrate
is $X$).  Repeatedly apply this operator until a fixed point is
reached.  The desired interpretations of the superstrate relations
constitute the least (with respect to componentwise inclusion) fixed
point of $\Gamma$.

\q
This definition reminds me of something else that I wanted to ask
you.  In \cite{dkal-abs}, you called this language ``liberal
Datalog,'' but, since you allow function symbols, it looks to me like
pure Prolog.  Isn't the presence or absence of function symbols the
essential difference between Prolog and (constraint) Datalog?

\a The intended semantics of Prolog uses an Herbrand universe, which
means a structure where every element is denoted by a unique ground
term.  The substrate structures of liberal Datalog are quite
arbitrary.  In particular, the functions of the structure need not be
free constructors.

\q So liberal Datalog is liberal even when compared to pure Prolog.

\a That's right.

Now let's see that the superstrate relations produced, over a
substrate $X$, by a liberal Datalog program can be defined in $X$ by
EFPL formulas in the sense of \cite{efp}.  In fact, we'll obtain the
required formulas in a simple, explicit manner from the given program
$\Pi$.

As a first step, we can rewrite $\Pi$ so that the head of each
rule involves no function symbols, i.e., each head looks like
$R(x_1,\dots,x_n)$ where the $x_i$ are variables.

\q
This was already explained in \cite{dkal} in the context of an
example, but the method is clearly general.  Given a rule of the form
$R(t_1,\dots,t_n)\leftarrow B$, where the $t_i$ are terms that need not
be variables, replace it with
\[
R(x_1,\dots, x_n)\leftarrow B\land\bigwedge_{i=1}^n(x_i=t_i)
\]
where the $x_i$ are distinct, fresh variables.  This modification of
$\Pi$ has no effect on the operator $\Gamma$ that it defines, so the
superstrate relations are unchanged.

\a
Right.  When making these modifications to $\Pi$, you can also
arrange that, if the same relation symbol $R$ occurs in the head of
several rules, then the same tuple of variables $x_1,\dots,x_n$ is
used for all these occurrences.

At this point, we'll gradually move from the syntax of logic rules to
the syntax of EFPL.  Specifically, we'll modify the rules of our
program some more, and the resulting bodies will no longer have the
form required in logic rules but rather will be EFPL formulas.

If several rules in the program begin with the same superstrate
symbol $R$ and therefore, by the preceding normalization, have the
same head $H$, then we combine these rules $H\leftarrow B_1$, \dots,
$H\leftarrow B_k$ into a single rule
\[
H\leftarrow (B_1\lor \dots \lor B_k).
\]

\q
This use of disjunction was allowed in \cite[Appendix~A.2.2]{dkal}.

\a
Yes, but there it was regarded as syntactic sugar, a mere abbreviation
of the $k$ separate rules $H\leftarrow B_j$.  Now, we want to regard
it as a single rule in its own right.

Next, if the body of a rule contains variables other than those in the
head, quantify them existentially.  That is, replace
$R(x_1,\dots,x_n)\leftarrow B$ with
\[
R(x_1,\dots,x_n)\leftarrow (\exists y_1)\dots(\exists y_r)\,B
\]
where $y_1,\dots,y_r$ are all the variables in $B$ other than
$x_1,\dots,x_n$.  Is it clear that this change doesn't affect the
superstrate relations?

\q
Yes.  In fact, it doesn't change the operator $\Gamma$ used to define
those relations.  The essential point is that the definition of
$\Gamma$ was already in terms of ``some assignment.''

\a Good.  Notice also that the bodies of our rules, after these
modifications, are still EFPL formulas.

\q
Sure.  In fact, they're in the existential fragment of first-order
logic.

\a Right.  Let $\Pi'$ be the current, modified program\footnote{To
  agree exactly with the syntax of \cite{efp}, $\Pi'$ should be
  regarded as a sequence of rules, rather than a set, by ordering its
  rules arbitrarily.  This pedantry is required because in \cite{efp}
  we defined the fixed-point construction ``LET \dots\ THEN \dots''
  using sequences between the LET and the THEN.  Sequences have the
  advantage that formulas are strings of symbols; sets would have the
  advantage of mathematical elegance, since the ordering in the
  sequence never matters.  The same comments apply to logic programs.
  It is curious that the directly writable, sequence convention is
  used in the mathematically oriented paper \cite{efp} while the more
  elegant, abstract, set convention is used in the
  application-oriented paper \cite{dkal}.}, let $R$ be a superstrate
relation, say $n$-ary, and consider the EFPL formula
$\phi(x_1,\dots,x_n)$ given by
\[
\text{LET }\Pi'\text{ THEN }R(x_1,\dots,x_n).
\]

\q
Are the variables $x_i$ after THEN the same ones that were used with
all occurrences of $R$ in head position in the program?

\a
They might as well be, but it doesn't matter.  All variables in the
$\Pi'$ part of $\phi$ are bound, either by the quantifiers that we
explicitly introduced into the bodies of rules or by the ``LET \dots\
THEN'' construction.  So it doesn't matter which variables they are.
The only reason we insisted on having the same variables for all
occurrences, in heads, of the same relation symbol is to be able to
combine those rules into a single rule.  Thus in $\Pi'$ each
superstrate relation symbol occurs in the head of exactly one rule, as
required by the syntax of EFPL.

\q
Wait a minute.  I see why each superstrate relation symbol occurs in
the head of at most one rule in $\Pi'$; if it was originally in more
than one, then you combined those rules using disjunction.  But why is
it in exactly one?  What if some superstrate symbol doesn't occur at
all?

\a
We ignored that situation because such a symbol would be interpreted
as the empty relation over any substrate, so there's really no point
in including it in the superstrate.  But, to be accurate, we should
cover this case as well, and the disjunction idea still works.  So if
the original program had no rules starting with the superstrate symbol
$R$, then $\Pi'$ would have one such rule, $R(x_1,\dots,x_n)\leftarrow
B$, where $B$ is the disjunction of no formulas (the bodies of all the
0 rules with $R$ in the head).  Since the disjunction of no formulas
is, by the only reasonable convention, \ttt{false}, we get the rule
$R(x_1,\dots,x_n)\leftarrow\ttt{false}$, which has the right semantical
effect.

\q
That's a pretty pedantic answer.

\a
It was a pretty pedantic question.

Coming back to the EFPL formula $\phi$ defined above, is it clear that
the relation it defines is exactly the interpretation of the
superstrate relation $R$  that is produced by the original logic
program $\Pi$?

\q Almost.  It's clear that, if we interpreted $\Pi'$, where it occurs
in $\phi$, in the same way that logic programs are interpreted, then it
would produce the same superstrate relations as $\Pi$.  But the
interpretation of rules, and specifically the operator $\Gamma$, is
not quite the same in logic programs as in EFPL.  In the semantics of
logic programs, the set of tuples described by a rule is \emph{added}
to the current interpretation of the relevant superstrate relation
symbol.  In the semantics of EFPL, the same set of tuples \emph{alone}
constitutes the new interpretation of that symbol.  In other words,
the $\Gamma$ operator for logic programs \cite{dkal} is explicitly
designed to be inflationary; that of EFPL \cite{efp} need not be
inflationary.

\a
That is true, but it doesn't matter.  A not-necessarily-inflationary
operator $\Delta$ and its explicitly inflationary variant $\Gamma$
defined by $\Gamma(A)=\Delta(A)\cup A$ have the same closed points.

\q What do you mean by closed points?

\a A closed point of $\Gamma$ is an $A$ such that $\Gamma(A)\subseteq
A$.  It's fairly common terminology to say that a set is closed under
an operator.  We say ``closed point'' rather than ``closed set''
because, when there are several superstrate predicate symbols, our
operators act on tuples of relations, not on single sets.

Coming back to the situation of an operator $\Delta$ and its
inflationary variant $\Gamma$, it's clear that they have the same
closed points.
For any monotone operator, the least fixed point
is also the least closed point. And our operators are monotone,
because superstrate relations occur only positively in the bodies of
logic programs (even after we modify the programs as above).  So
$\Gamma$ and $\Delta$ have the same least fixed point.

\q OK.  Actually, I now see another reason why we can ignore the
inflationary aspect of the $\Gamma$ in \cite{dkal}.  If we think of
the least fixed point of a monotone operator $\Delta$ as being
constructed by iterating $\Delta$, then the sequence of iterates
is non-decreasing.  So, for every $A$ in this sequence,
$\Delta(A)=\Gamma(A)$.

\a Right.  So this completes the proof that the superstrate relations
defined, over a given substrate structure $X$, by a logic program
$\Pi$, as in \cite{dkal}, are also defined over $X$ by EFPL formulas
$\phi$.  Furthermore, the transformation of $\Pi$ into $\phi$ is
uniform, i.e., the same for all $X$.

\q
Yes.  In fact it proves a bit more, namely that any logic program can
be translated into EFPL formulas of a rather special form: A single
LET \dots\ THEN construct, where the part after THEN is an atomic
formula consisting of a relation symbol followed by variables.
Furthermore, the bodies of the inductive definitions (rules) after LET
are existential first-order formulas.

That worries me, because for the other half of the equivalence between
logic programs and EFPL, you'll have to find logic programs equivalent
to arbitrary EFPL formulas, not just those of this special form.

\a
That's right, but you needn't worry.  Every EFPL formula is equivalent
to one in this special form, and the equivalence to logic programs is
one way to prove this.

\section{From Formulas to Logic Programs}

\a Given an arbitrary EFPL formula $\phi(u_1,\dots,u_n)$ with free
variables among those indicated, we shall transform it into a logic
program $\Pi$ such that a particular $n$-ary superstrate relation, as
defined by $\Pi$ over any substrate $X$, is the set of $n$-tuples that
satisfy $\phi$ in $X$.  That will complete the proof that logic
programs and EFPL formulas are equivalent; they can be regarded as two
ways of presenting the same logic.

As a first step, we'll show that every EFPL formula is logically
equivalent to one of the special form
\[
\text{LET }P_1(\vec x^1)\leftarrow\delta_1,\dots,
P_k(\vec x^k)\leftarrow\delta_k\text{ THEN }\psi
\]
where all the formulas $\delta_i$ and $\psi$ are existential
first-order formulas and where the free variables of any $\delta_i$
are among the variables $\vec x^i$ serving as the arguments of the
corresponding $P_i$ in the definition $P_i(\vec
x^i)\leftarrow\delta_i$.

\q Since ``LET \dots\ THEN'' binds these $\vec x^i$s,
the only free variables in such a formula are those in $\psi$.

\a Right; that will simplify part of the proof.  We should also
mention that $k=0$ is allowed; then the formula above amounts to just
$\psi$.

\q
I bet your proof that all EFPL formulas are equivalent to ones of this
special form is an induction on formulas, and by allowing
$k=0$ you've made the cases of atomic formulas and of negated atomic
formulas trivial (whereas otherwise they would only have been
obvious).

\a Right on both counts.  Conjunction and disjunction are also easy.
Given two formulas in the desired form, rename the bound predicate
variables of the LET-clause --- the $P_i$'s in the notation above ---
in one of them so as to be distinct from those of the other formula.
Then just combine their LET-clauses and form the conjunction or
disjunction of the THEN-clauses.

Existential quantification is even easier.  Leave the LET-clause alone
and quantify the THEN-clause.

\q
This simple argument for the existential quantifier case makes use of
your convention that the only free variables in $\delta_i$ are
among the $\vec x^i$.  Without this convention, you'd have to consider
the possibility that some other variable free in some $\delta_i$ is
being quantified.

But, in keeping with the principle that there's no free lunch, it seems
that you'll have to pay for this convention in the one remaining case
of your induction.  Given a LET-THEN formula, you'll have to get rid
of any extraneous free variables in its LET-clause.

\a
You're right, but in this case the lunch is fairly cheap.

Suppose we're given a formula $\text{LET }P(x)\leftarrow\delta\text{
  THEN }\psi$ where both $\delta$ and $\psi$ are of the desired form.

\q
Wait a minute.  Are you assuming that there's only one constituent
$P(x)\leftarrow\delta$ in the LET part and that $P$ is unary?

\a Yes, but this is only for notational simplicity.\footnote{It is
  also known that, at least in the presence of two constant symbols,
  simultaneous positive recursions can be reduced to a single positive
  recursion, albeit for a relation of higher arity.  See
  \cite[Theorem~1C.1]{ynm}.  The context in \cite{ynm} is positive
  recursion over full first-order logic, but universal quantification
  isn't used for this theorem.} The general case would involve a lot of
subscripts but no new ideas.  Notice, in particular, that if we have
several $P$'s with their corresponding $\delta$'s, and if each
$\delta$ has a LET-clause with several consitituents, defining
predicates $Q$, then each of those $Q$'s needs two subscripts
--- the first to
tell which $\delta$ it's in and the second to tell where it is in that
$\delta$'s LET-clause --- and the range of the second subscript
depends on the first.  Subscript-juggling in such a case tends to
obscure the proof.

\q
OK, go ahead with your ``one unary predicate'' proof.  Maybe
afterward I'll figure out all the subscripts for the general case on
my own.

\a
Let's start by taking our formula,
\[
\text{LET }P(x)\leftarrow\delta\text{ THEN }\psi,
\]
and showing how to eliminate any extraneous free variables from
$\delta$.  Continuing to avoid uninformative subscripts, let's suppose
$y$ is the only variable other than $x$ that is free in $\delta$.
Then we claim our formula is equivalent to
\[
\text{LET }P'(x,y)\leftarrow\delta'\text{ THEN }\psi',
\]
where $\delta'$ and $\psi'$ are obtained from $\delta$ and $\psi$ by
replacing each atomic subformula of the form $P(t)$ with $P'(t,y)$.
(Of course, we assume that bound variables in $\delta$ and $\psi$ have
been renamed if necessary so that the $y$'s introduced here don't
become accidentally bound.)

To see that the new formula is equivalent to the original, consider
the binary relation obtained as (the interpretation of) $P'$ from the
recursion $P'(x,y)\leftarrow\delta'$.  If you fix any particular value
$b\in X$ for $y$, then the resulting unary relation, $P'(x,b)$ is
exactly the relation defined by the original clause,
$P(x)\leftarrow\delta$ with $y$ assigned the value $b$.

\q
Yes, that's easy to see if you think of the iterative process leading
to the fixed points that interpret $P$ and $P'$.  In the new clause,
$P'(x,y)\leftarrow\delta'$, $y$ behaves simply as a parameter.  So, as the
binary relation defined by this clause grows, from $\emp$ toward the
fixed point $P'$, its unary section obtained by fixing the second
argument as $b$ grows exactly according to the original clause
$P(x)\leftarrow\delta$ with $y$ denoting $b$.  In particular, the
agreement between $P$ and a section of $P'$ occurs not only for the
final fixed points but stage by stage during the iteration.

\a That's right.  But one can also verify the final agreement directly
in terms of least fixed points without referring to the iteration.  If
$P'$ is the least fixed point of the new iteration, then each of its
sections, say at $b$, is the least fixed point of the old iteration
with $y$ denoting $b$.

\q
I see that the section is a fixed point of the old operator, simply
because $P'$ itself is a fixed point of the new one, but why is it the
least fixed point?

\a If you had a smaller fixed point $P^-$ for the old operator, then
you could replace the $b$ section of $P'$ with $P^-$ while leaving all
the other sections unchanged.  The result would be a smaller fixed
point than $P'$ for the new operator.  The point here is that we can
modify a single section independently of the others because the new
operator works on each section separately.

\q
I see; this is what I expressed earlier by saying that $y$ behaves
simply as a parameter.

OK, so you can get rid of extraneous free variables in $\delta$.  But
your unary $P$ has become a binary $P'$.

\a
It's still the case that higher arities (or more $P$'s) contribute
only notational complications.  So, if you don't mind, we'll revert to
the unary notation and we'll drop the primes on $P$, $\delta$, and
$\sigma$.  In other words, we'll return to the original notation
$\text{LET }P(x)\leftarrow\delta\text{ THEN }\psi$ but with the
assumption that only $x$ is free in $\delta$.

\q
OK; I never was a big fan of subscripts.

\a
Good. So let's consider this formula $\phi$:
\[
\text{LET }P(x)\leftarrow\delta\text{ THEN }\psi.
\]
By induction hypothesis, we know that $\delta$ and $\psi$ are already
of the desired form, say $\delta$ is
\[
\text{LET }R(y)\leftarrow\rho\text{ THEN }\pi,
\]
and $\psi$ is
\[
\text{LET }S(z)\leftarrow\sigma\text{ THEN }\theta,
\]
where $\rho,\pi,\sigma, \theta$ are existential first-order formulas,
and their free variables are among those indicated here:
\[
\rho(y),\quad\pi(x),\quad\sigma(z),\quad\theta(u).
\]

\q
What's this $u$?  It wasn't in any of the previous formulas.

\a $u$ represents whatever variables are free in the whole formula
$\phi$.  They were called $u_1,\dots,u_n$ at the beginning of this
half of the proof, but, as usual, we now pretend, for notational
simplicity, that there's only one such variable.

\q OK.  All your other limitations on free variables are based on the
fact that, by induction hypothesis and by the preceding discussion,
none of the three LET-clauses have extraneous free variables.  In
particular, any variable free in $\pi$ would also be free in $\delta$
and therefore can only be $x$.

\a
That's right.  Let's write out $\phi$ in detail, exhibiting not only
the free variables in each part (as above) but also the predicate
variables, $P,R,S$, that could occur in each part.  So $\phi$ looks
like
\begin{multline*}
\text{LET }P(x)\leftarrow[\text{LET }R(y)\leftarrow\rho(P,R,y)
\text{ THEN }\pi(P,R,x)]\\
\text{THEN } [\text{LET }S(z)\leftarrow\sigma(P,S,z)\text{ THEN
}\theta(P,S,u)].
\end{multline*}

\q Please wait a minute while I check your claims about which
predicates can occur where. \dots\ OK, I agree with what you wrote.
The point is that the predicate variable introduced before a
$\leftarrow$ in a LET-clause is allowed to occur at the right of that
$\leftarrow$ in that LET-clause and also in the associated THEN-clause
but not elsewhere.

\a
Right.  Now we claim that $\phi$ is equivalent to the following
formula $\phi'$:
\begin{multline*}
  \text{LET }P(x)\leftarrow\pi(P,R,x),\ R(y)\leftarrow\rho(P,R,y),\
  S(z)\leftarrow\sigma(P,S,z)\\
  \text{THEN }\theta(P,S,u).
\end{multline*}

\q Essentially, you've just lumped together all the LET-clauses in
$\phi$, ignoring the nesting of the clause for $R$ inside the clause
for $P$, and made one big LET-clause out of all of them.  Not very
subtle.

\a But it works.  The first step toward the proof that it works is
setting up some notation that is neither cumbersome nor ambiguous.
(Either problem alone is easily avoided.)  We propose the following.

Fix a structure and a value for the free variable $u$ of $\phi$.
Since they're fixed, we won't mention them in our notation.  To
further simplify the notation, we'll generally use the same symbols
for syntactic entities (like the predicates $P,R,S$ and the variables
$x,y,z$) and possible semantic interpretations of them in our
structure.

Now let's look at our formula $\phi$ and set up some notation for the
various fixed points that occur in it.  We begin with the LET-clause
$R(y)\leftarrow\rho(P,R,y)$ that defines the fixed-point
interpretation for $R$.

\q
Why not start with the first LET-clause, the one for $P$?

\a The defining formula $\delta$ in that clause involves the fixed
point for the $R$ clause, so it's useful to settle the $R$ part of the
notation first.

For any particular interpretation of the predicate $P$ --- and, as
indicated above, we'll use the same symbol $P$ for the interpretation
--- $\rho$ defines a least fixed point that we'll call $R^\infty(P)$.

Next, the LET-clause for $P$ amounts to using the definition
$\pi(P,R,x)$ but with $R$ interpreted as $R^\infty(P)$.  (Indeed,
that's the semantics of ``LET $R(y)\leftarrow\rho(P,R,y)$ THEN
$\pi(P,R,x)$.'') Note carefully that the monotone operator described
by this clause,
\[
P\mapsto\{x:\pi(P,R^\infty(P),x)\}
\]
uses its input $P$ twice --- first in the first argument of $\pi$ and
again via the dependence on $P$ of the second argument $R^\infty(P)$.
We denote the least fixed point of this operator by $P^\infty$.

Similarly, the LET-clause for $S$ uses the definition $\sigma(P,S,z)$
with $P$ interpreted as $P^\infty$.  We write $S^\infty$ for the least
fixed point of this operator.

Next, we need notation for the three predicates obtained as the least
fixed point of the simultaneous recursion in $\phi'$.  Having already
used superscripts $\infty$ for another purpose, we'll use stars
instead for this triple of fixed points, calling them $P^*,Q^*,R^*$.

In connection with all these fixed points, it is useful to remember
that the least fixed point is also the least closed point.  Thus, for
example, $P^*,Q^*,R^*$ can be characterized as the smallest relations
that simultaneously satisfy the implications
\begin{align}
  \forall x\,&[\pi(P^*,R^*,x)\implies P^*(x)]\\
  \forall y\,&[\rho(P^*,R^*,y)\implies R^*(y)]\\
  \forall z\,&[\sigma(P^*,S^*,z)\implies S^*(z)]
\end{align}
(whereas for fixed points we'd write bi-implications).

The essence of the proof is to show that the fixed points arising from
$\phi$ and from $\phi'$ match as follows.
\[
P^*=P^\infty,\quad R^*=R^\infty(P^*), \quad\text{and}\quad
S^*=S^\infty.
\]

\q It's clear that, once you establish these equations, you're done.
After all the truth value of $\phi$ is, by definition, obtained by
evaluating $\theta$ with $P$ and $S$ interpreted as $P^\infty$ and
$S^\infty$, while the truth value of $\phi'$ is obtained by evaluating
the same $\theta$ (in the same structure with the same value for $u$,
as fixed earlier) with $P$ and $S$ interpreted as $P^*$ and $S^*$.  In
fact, this last part of the proof won't need the equation for $R$, with
its curious mixture of $\infty$ and $*$ on the right side; presumably
that equation is needed as an intermediate step in the proof of
the other two equations.

So please go ahead and prove the $*=\infty$ equations.

\a OK.  We'll start by showing that $R^*=R^\infty(P^*)$.  Formula~(2)
says exactly that $R^*$ is a closed point of the operator
$R\mapsto\{y:\rho(P^*,R,y)\}$.  According to the definition of
$R^\infty(P)$, applied with $P$ instantiated as $P^*$, the least
closed point of this operator is $R^\infty(P^*)$.  So we immediately
have that $R^\infty(P^*)\subseteq R^*$.

Furthermore, all three of the formulas~(1), (2), and (3) remain true
if we replace $R^*$ by $R^\infty(P^*)$.  For (1), this follows from
the fact that $R$ is a positive predicate symbol (otherwise it
couldn't have occurred on the left of $\leftarrow$ in $\phi$) and so
$\pi$ is monotone with respect to $R$.  When we replace $R^*$ by
$R^\infty(P^*)$, the interpretation of $R$ can only decrease.  That
strengthens the antecedent in (1) and thus preserves the truth of (1).
The argument for (2) is easier; the result of replacing $R^*$ by
$R^\infty(P^*)$ there is just the fact that $R^\infty(P^*)$ is by
definition closed under the operator defined by $\pi$ with $P$
interpreted as $P^*$.  Finally, the case of (3) is trivial, as $R^*$
doesn't occur there.

Thus, the triple $P^*,R^\infty(P^*),S^*$ is closed under the
simultaneous recursive operator whose least closed point is
$P^*,R^*,S^*$.  Therefore $R^*\subseteq R^\infty(P^*)$, and the
$*=\infty$ equation for $R$ is proved.

Let's turn next to the equation for $P$.  In view of what we've
already proved, formula~(1) can be written as
\[
\forall x\,[\pi(P^*,R^\infty(P^*),x)\implies P^*(x)],
\]
which says that $P^*$ is closed under the operator
$P\mapsto\{x:\pi(P,R^\infty(P),x)\}$, whose least closed
point is $P^\infty$.  So we have $P^\infty\subseteq P^*$.

Furthermore, all three of the formulas~(1), (2), and (3) remain true
if we replace $P^*$ by $P^\infty$ and replace $R^*$ by
$R^\infty(P^\infty)$.  In the case of (1), this is just the closure
condition defining $P^\infty$.  In the case of (2), it's the closure
condition defining $R^\infty(P^\infty)$.  And in the case of (3), it
follows from the facts that $\sigma$ is monotone with respect to $P$
and $P^\infty\subseteq P^*$.

Thus, the triple $P^\infty, R^\infty(P^\infty),S^*$ is closed under
the simultaneous recursive operator whose least closed point is
$P^*,R^*,S^*$.  Therefore $P^*\subseteq P^\infty$, and the $*=\infty$
equation for $P$ is proved.

Finally, we turn to $S$.  In view of the equations already proved,
formula~(3) is equivalent to
\[
\forall z\,[\sigma(P^\infty, S^*,z)\implies S^*(z)],
\]
which says that $S^*$ is closed under the operator whose least closed
point is $S^\infty$.  Therefore, $S^\infty\subseteq S^*$.

Furthermore, all three of the formulas~(1), (2), and (3) remain true
if we replace $P^*$ by $P^\infty$, $R^*$ by $R^\infty(P^\infty)$, and
$S^*$ by $S^\infty$.  Since $S^*$ doesn't occur in (1) and (2), the
argument given for them above still applies.  As for (3), the formula
we get is just the closure requirement in the definition of
$S^\infty$.

Thus, the triple $P^\infty, R^\infty(P^\infty),S^\infty$ is closed under
the simultaneous recursive operator whose least closed point is
$P^*,R^*,S^*$.  Therefore, $S^*\subseteq S^\infty$, and so the proof
is complete.

\q You mean the proof of the $*=\infty$ equations. So you've completed
the inductive proof that every EFPL formula can be put into the normal
form LET $P(x)\leftarrow\delta$ THEN $\psi$, with $\delta$ and $\psi$
in existential first-order logic.  And you have the additional
normalization that $\delta$ has no free variables but $x$ (except
that you might really have lots of $P$'s of arbitrary arities).  You
still have to convert this into a logic program of the form used in
\cite{dkal}.

\a Right, but the rest is fairly easy.

First, we can arrange that the formula $\psi$ after ``THEN'' is an
atomic formula $Q(\vec u)$ where $\vec u$ is a tuple of distinct
variables (not compound terms).  Indeed, if we let $\vec u$ list all
the variables free in the given formula LET $P(x)\leftarrow\delta$
THEN $\psi$ (hence free in $\psi$) and if we let $Q$ be a new,
positive predicate symbol of the right arity, then this given formula
is easily equivalent to
\[
\text{LET }P(x)\leftarrow\delta,\ Q(\vec u)\leftarrow\psi
\text{ THEN }Q(\vec u).
\]

\q I see: Since $Q$ doesn't occur in $\psi$, the ``recursive''
definition $Q(\vec u)\leftarrow\psi$ isn't really recursive.  The
relevant iteration takes just one step (once $P$ has reached its fixed
point) and, in effect, makes $Q(\vec u)$ an alias for $\psi$.

\a
Right.  Furthermore, as $Q$ doesn't occur in $\delta$, nothing has
changed in the recursive definition of $P$.

So our formula has been equivalently rewritten as
\[
\text{LET }P_1(\vec x^1)\leftarrow\delta_1,\ \dots,\
P_k(\vec x^k)\leftarrow\delta_k\text{ THEN }P_k(\vec u).
\]

\q
You must really be getting near the end of the proof, since you've
restored the multiple $P$'s and their subscripts.

\a Yes.  What remains is to transform each of the existential
first-order formulas $\delta_i$ as follows.

We can put each $\delta_i$ into prenex form and then put its
quantifier-free matrix into disjunctive normal form.  So $\delta_i$
now looks like
\[
\exists \vec y^i\ \bigvee_r\bigwedge_s\alpha_{i,r,s}(\vec x^i,\vec y^i),
\]
where the $\alpha$'s are atomic or negated atomic formulas.

Then the required logic program $\Pi$ consists of the rules
\[
P_i(\vec x^i)\leftarrow \bigwedge_s\alpha_{i,r,s}(\vec x^i,\vec y^i),
\]
one for each $i$ and $r$.

\q
This is very similar to what happened in the first part of the proof,
translating logic programs into EFPL formulas.  The monotone operators
defined by the transformed EFPL formula and by the logic program are
identical as operators.  So they certainly have the same least fixed
point.  In particular, the $P_k$ component of that least fixed point,
which is one of the superstrate relations defined by the program, is
the interpretation, in the substrate, of the original EFPL formula.

\begin{rmk}
  Although, as mentioned in a footnote earlier, \cite{dkal} is directed
  toward applications and \cite{efp} is more theoretical, EFPL does
  have one advantage over liberal Datalog from the programming point
  of view.  In a liberal Datalog program, all variables are global,
  but in an EFPL formula, the LET-THEN construction provides local
  variables with scopes.  The latter can be important for large-scale
  programming, by making it easy to assemble small modules into a
  large program (or formula).
\end{rmk}

\subsection*{Acknowledgment} We thank Nikolaj Bj\o rner for helpful
comments on a draft of this paper.

\end{document}